\documentclass{article}
\def\real{{\rm I \kern-0.2em R}}       
\newcommand{\R}{{\rm{I\!\!\!R}}}

\newcommand{\bbR}{\overline{\R}_+}

\input{psfig.sty}
\begin{document}

\title{The Generalized Riemann or Henstock Integral Underpinning 
Multivariate Data Analysis: Application to Faint Structure
Finding in Price Processes}

\author{Pat Muldowney (1) and Fionn Murtagh (2, *) \\
(1) School of International Business \\
University of Ulster, Magee College \\
Londonderry BT48 7JL, Northern Ireland \\
Email p.muldowney@ulster.ac.uk \\
(2) Department of Computer Science \\
Royal Holloway, University of London \\
Egham, Surrey TW20 0EX, England \\
Email fmurtagh at acm dot org \\
$\ast$ Author for correspondence}

\maketitle

\begin{abstract}
Practical data analysis involves many implicit or explicit 
assumptions about the good
behavior of the data, and excludes consideration of various potentially
pathological or limit cases.  In this work, we present a new general theory 
of data, and of data processing, to bypass some of these assumptions.  The new 
framework presented is focused on integration, and has direct
applicability to expectation, distance, correlation, and aggregation.  
In a case study,
we seek to reveal faint structure in financial data.  Our new foundation 
for data encoding and handling offers increased justification for our 
conclusions.   
\end{abstract}

\noindent
{\bf Keywords:} data coding, data encoding, data valuation, 
correspondence analysis, 
hierarchical clustering, 
geometric Brownian
motion, financial modeling, time series prediction, data aggregation

\section{Introduction}

We develop a theory of data for contingency table data analysis, a 
priority area of application of correspondence analysis.  Much of the
foundations of data theory that we discuss are quite general to data 
analysis, and independent of the correspondence
analysis. Motivation includes the following.

Correspondence analysis is carried out on a cloud of points (rows, 
columns) through finding of principal directions of elongation, etc.  What
legitimizes our assumption of a compact cloud of points?
More generally, what legitimizes our data analysis of a given data set,
when we assume that the data set is a sampling of facets or events (which 
are to be explained and interpreted through the data analysis)?   
Should we instead allow for singularities or other pathologies or
irregularities in such a cloud of points?  The data analyst, in a 
somewhat slipshod approach to analyzing data, ignores such issues, 
and instead cavalierly takes data as sometimes discrete and sometimes
continuous.
As an example of such singularities, consider the preprocessing of
data using normalization through taking the logarithm (common in dealing 
with astronomical stellar magnitudes, or financial ratios).  
Such normalization can potentially give 
rise to undefined data values.  Why do we consider that our input data sets
do not also contain undefined data values?
In all generality, what justifies the ruling out of 
such pathologies in our input data?

The number of attributes used to characterize our observations is 
possibly infinite.  Can our general foundations cope with this?  A priori 
the answer is clearly no.  In this article, we describe a foundation
for data analysis, 
based on Henstock's approach to integration, which allows us to bypass
such pitfalls in a rigorous manner.  

We need a theory which begins with empirical distribution functions 
deduced from empirical data (i) for which there is no analytical description,
and (ii) that are amenable to empirical computation.

We propose in this article a foundation for data analysis which is 
at the level of the data, rather than at higher levels of
 model fitting, so that we are 
fully compatible thereafter with all statistical modeling approaches.  
In passing we will note how quantitative and qualititive data coding
are encompassed within our approach (in section \ref{sectGenRie}).  Neither
can be considered as the more legitimate.
There is no one necessary a priori statistical model to be used
because 
there is no one necessary a priori morphology for a data cloud.  
(See section \ref{cylint}.)
Nor is there any one necessary level of resolution in data encoding
(section \ref{infint}).
Empirical distribution functions can be deduced from
empirical data for which there is no analytical
description; and then the Riemann sums, with their finite number
of terms, are amenable to empirical computation.

In multivariate data analysis, the input data set is assumed to be
representative and comprehensive.  However the former cannot do justice to 
an unknown (and perhaps unknowable) underlying (physical, social, etc.) 
reality.  The latter is approximated very crudely in practice.  Can these
goals of representativity and comprehensiveness even hypothetically be well
approximated in practice?  Only with the framework that we present in this 
article can pathologies be excluded (in regard to representativity), and (in 
regard to comprehensiveness) can we be at ease with infinite dimensional
spaces.  

As is clear from this list of motivations, we are concerned with the 
well-foundedness of numerical data, which will subsequently be subject to a
statistical data analysis.  
The supposition that (multivariate, time series, etc.) data can be 
addressed as such has only been examined in terms of measurement theory 
(ordinal, interval, qualitative, quantitative, etc.)
 or levels of measurement by S.S. Stevens 
in the 1940s (see Velleman and Wilkinson, 1984).  
However  suppositions regarding input data have not been examined 
before in terms of the data set giving rise a 
well-behaved and exploitable processing input.  We will do so in this article
by showing how the Henstock or generalized Riemann theory of integration 
also provides a basis for asserting: {\em a numerical data set can be
analyzed}.   The focus on integration, and the perspective introduced,
 is easily extended to expectation, scalar product, distance, correlation,
data aggregation, and so on.  

A word on terminology used here: all statistical analysis of data starts
with (qualitative or quantitative) data in numeric form, presupposing a 
valuation function mapping facets (or events) of the domain studied onto 
numerical values.  We speak of this as data valuation, or more usually in 
this context as data encoding.  The bigger picture of data encoding together
with data normalization or other preprocessing, or indeed processing 
in the data analysis pipeline, is referred to in this article as data 
coding.  

\section{Integration Background}

Probability theory, with foundations provided by Kolmogorov, is 
based on probability measures on 
algebras of events and based ultimately on the Lebesgue 
integral.  
Lebesgue's just happened to be  the first of a number of such
investigations into the nature of mathematical integration during
the twentieth century.

Subsequent developments in integration, by Perron, Denjoy,
Henstock and Kurzweil, have similar properties and were devised to
overcome shortcomings in the Lebesgue theory. See Gordon (1994) 
for detailed comparison of modern theories of integration.
However, theorists of probability and random variation have not
yet really ``noticed'', or taken account of, these developments in
the underlying concepts. There are many benefits to be reaped by
bringing these fundamental new insights in integration or
averaging to the study of random variation, and this article aims
to demonstrate some of them in the context of data coding.

It is possible to formulate a theory of random variation
and probability, linked to data coding, on the basis of a conceptually simpler
Riemann-type approach, and without reference to the more difficult
theories of measure and Lebesgue integration. 

In particular it is possible to present a Riemann-type model of
data encoding in which a valuation (potentially a data value) 
is a limit
of Riemann sums formed by suitably partitioning the sample space
$\Omega$ in which the process $x$ takes its values. See Muldowney (1999, 
2000/2001).

To contrast (traditional) Legesgue and (more recent) Riemann integration,
consider determining a mean value. Suppose the sample space is
the set of real numbers, or a subset of them. If successive
instances of the random variable are obtained, we might partition
the resulting data into an appropriate number of classes; then
select a representative value of the random variable from each
class; multiply each of the representatives by the relative
frequency of the class in which it occurs; and add up the
products. The result is an estimate of the mean value of the
random variable. Table \ref{riem1} illustrates
this procedure. The sample space is partitioned into intervals
$I^{(j)}$ of the sample variable $x$, the random variable is
$f(x)$, and the relative frequency of the class $I^{(j)}$ is
$F(I^{(j)})$.

\begin{table}
\begin{displaymath} 
\begin{array}{lll}
\mbox{Interval}  & \mbox{Random variable}  & \mbox{Relative frequency} \\ 
 \hline
  I^{(1)} & f(x^{(1)}) & F(I^{(1)}) \\
I^{(2)} & f(x^{(2)}) & F(I^{(2)}) \\
 \vdots & \vdots & \vdots\\
 I^{(n)} & f(x^{(n)}) & F(I^{(n)})
\end{array}
\end{displaymath}
\caption{For each $j$, the number $x^{(j)}$ is a representative element
selected from $I^{(j)}$ or its closure. The resulting estimate of
the mean value of the random variable $f(x)$ is $ \sum_{j=1}^n
f(x^{(j)}) F(I^{(j)})$.}
\label{riem1}
\end{table}

The approach to  random variation 
that we are concerned with in this article consists of a
formalization of this relatively simple Riemann sum technique
which puts at our disposal powerful results in analysis such as
the Dominated Convergence Theorem.

In contrast the Kolmogorov approach requires, as a preliminary, an
excursion into abstract measurable subsets $A_j$ of the sample
space, $\Omega$ (Table \ref{kol1}).

\begin{table}
\begin{displaymath}
\begin{array}{lll}
x&f(x)&P \\ \hline A^1 & y^1 = f(x^{(1)}) & P(A^1) \\
A^2 & y^2 = f(x^{(2)}) & P(A^2) \\
 \vdots &\vdots &\vdots \\
 A^n & y^n = f(x^{(n)}) & P(A^n)
\end{array}
\end{displaymath}
\caption{Here, $x$ is again a representative member of a sample space
$\Omega$ which corresponds to the various potential occurrences or
states in the ``real world'' in which measurements or observations
are taking place on a variable whose values are unpredictable and
which can only be estimated beforehand to within a degree of
likelihood. A probability measure $P$ is posited on a sigma-algebra of
events $A$.}
\label{kol1}
\end{table}

In practice, $\Omega$ is often identified with the
real numbers or some proper subset of them; or with a Cartesian
product, finite or infinite, of such sets. In Table \ref{kol1},
numbers $y^j$ are chosen in the range of values of the random
variable $f(x)$, and $A^j$ is $f^{-1}([y^{j-1},y^j[)$. The
resulting $\sum_{j=1}^n y^j P(A^j)$ is an estimate of the expected
value of the random variable $f(x)$. But the $P$-measurable sets
$A^j$ are mathematically abstruse, and they can place heavy
demands on the understanding and intuition of anyone who is not
well-versed in mathematical analysis. For instance, it can be
difficult for a non-specialist to visualize a measurable set $A$ in terms
of laboratory, industrial or financial measurements of some
real-world quantity.

In contrast, the data classes $I^{(j)}$ of elementary statistics
in Table \ref{riem1} are easily understood as real intervals, of one
or more dimensions; and these are the basis of the Riemann
approach to random variation.

To illustrate the Lebesgue-Kolmogorov approach, 
suppose $X$ is a normally distributed random
variable in a sample space $\Omega$. Then we can represent
$\Omega$ as $\R$, the set of real numbers; with $X$ represented as
the identity mapping $X:\R \rightarrow \R$, $X(x)=x$; and with
distribution function $F_X$ defined on the family
$\mathcal{I}_{\R}$ of intervals $I$ of $\R$, $F_X:
\mathcal{I}_{\R} \rightarrow [0,1]$:
\begin{equation}\label{normal}
F_X(I) = \frac 1{\sqrt{2\pi}} \int_I e^{-s^2} ds.
\end{equation}
Then, in the Lebesgue-Kolmogorov approach, we generate, from the
distribution function $F_X$, a probability measure $P_X:
\mathcal{A}_{\R} \rightarrow [0,1]$ on the family
$\mathcal{A}_{\R}$ of Lebesgue measurable subsets of $\Omega=\R$.
So the expectation $E^P(f)$ of any $P_X$-measurable function $f$
of $x$ is the Lebesgue integral $\int_\Omega f(x) dP_X$. With
$\Omega$ identified as $\R$, this is just the Lebesgue-Stieltjes
integral $\int_{\R} f(x) dF_X$, and, since $x$ is just the
standard normal variable of (\ref{normal}), the latter integral
reduces to the Riemann-Stieltjes integral -- with Cauchy or
improper extensions, since the domain of integration is the
unbounded $\R = ]-\infty,\infty[$.

In presenting this outline we have skipped over many steps, the
principal ones being the probability calculus and the construction
of the probability measure $P$. It is precisely these steps which
cease to be necessary preliminaries if we take a generalized
Riemann approach, instead of the Lebesgue-Kolmogorov one, in the
study of random variation.

Because the generalized Riemann approach does not make use of an
abstract measurable space $\Omega$ as the sample space, from here
onwards we will take as given the identification of the sample
space with  $\R$ or some subset of $\R$, or with a Cartesian
product of such sets, and take the symbol $\Omega$ as denoting
such a space. Accordingly we will drop the traditional notations
$X$ and $f(X)$ for denoting  random variables. Instead a random
variable will be denoted by the variable (though unpredictable)
element $x$ of the (now Cartesian) sample space, or by some
function $f(x)$ of $x$. The associated likelihoods or
probabilities will be given by a distribution function $F(I)$
defined on intervals (which may be Cartesian products of
one-dimensional intervals) of $\Omega$. Whenever it is necessary
to relate the distribution function $F$ to its underlying random
variable $x$, we may write $F$ as $F_x$.

\section{A Generalized Riemann Approach: From Distribution Functions 
Rather Than From Probability Measures}
\label{sectGenRie}

The standard approach starts with a probability measure $P$
defined on a sigma-algebra of measurable sets in an abstract
sample space $\Omega$; it then deduces probability density
functions $F$. These distribution functions (and not some abstract
probability measure) are the practical starting point for the
analysis of many actual random variables -- normal (as described
above in (\ref{normal})), exponential,
 Brownian, geometric Brownian, and so on, i.e.\ practical data analysis.

In contrast, the generalized Riemann approach posits the
probability distribution function $F$ as the starting point of the
theory, and proceeds along the lines of the simpler and more
familiar (Table \ref{riem1}) instead of the more complicated and less
intuitive (Table \ref{kol1}).

To formalize the concepts, a random variable (or
\emph{observable}) is now taken to be a function $f(x)$ defined on
a domain $\Omega= S^B= \prod \{S:B\} $ where $S$ is $\R$ or some
subset of $\R$ and $B$ is an indexing set which may be finite or
infinite; the elements of $\Omega$ being denoted by $x$; along
with a likelihood function $F$ defined on the intervals of $\prod
\{S:B\}$.

In some basic examples such as throwing dice, $S$ may be a set
such as $\{1,2,3,4,5,6\}$, or, where there is repeated sampling, a
Cartesian product of such sets. Alternatively, $S$ will be the set of positive
numbers $\R_+$.  So quantitative and qualitative data encoding are easily 
supported.  

The Lebesgue-Kolmogorov approach develops probability density
functions $F$ from probability measures $P(A)$ of measurable sets
$A$. Even though distribution functions are often the starting
point in practice (as in (\ref{normal}) above), Kolmogorov gives
primacy to the probability measures $P$, and they are the basis of
the calculus of probabilities, including the crucial relation
\begin{equation}\label{measure}
P(\cup_{j=1}^\infty A_j) = \sum_{j=1}^\infty P(A_j).
\label{eqn2}
\end{equation}
 Viewed as an axiom, relation (\ref{eqn2}) is a somewhat mysterious
statement about rather mysterious objects. But it is the lynch-pin
of the Lebesgue-Kolmogorov theory, and without it the twentieth
century understanding of random variation would have been
impossible.

The generalized Riemann approach starts with probability density
functions $F_x$ defined only on intervals $I$ of the sample space
$\Omega=S^B$ . We can, as shown below (\ref{prob}), deduce from
this approach probability functions $P_x$ defined on a broader
class of ``integrable'' sets $A$, and a calculus of probabilities
which includes the relation (\ref{measure})---but as a theorem
rather than an axiom.

What, if any, is the relationship between these two approaches to
random variation? There is a theorem (Muldowney and Skvortsov, 2001/2002) 
which states that
every Lebesgue integrable function (in $\R^B$) is also generalized
Riemann integrable. In effect, this guarantees that every result
in the Lebesgue-Kolmogorov theory also holds in the generalized
Riemann approach. So, in this sense, the former is a special case
of the latter.

The key point in developing a rigorous theory of random variation
(which supports data valuation and hence data analysis) 
 by means of generalized Riemann integration is,  following the
 scheme of  Table \ref{riem1}, to partition the domain
 or sample space $\Omega=S^B$, in an
appropriate way, as we shall proceed to show. (Whereas in the
Lebesgue-Kolmogorov-It\^{o} approach we step back from
Table \ref{riem1}, and instead use Table \ref{kol1} supported by
(\ref{measure}). The two approaches part company at the
Tables \ref{riem1} and \ref{kol1} stage.)

In the generalized Riemann approach we focus on the classification
of the sample data into mutually exclusive classes or intervals
$I$.  I.e., through data encoding we undertake 
partitioning of the sample space $\Omega= S^B$ into mutually
exclusive intervals $I$.

In pursuing a rigorous theory of random variation along these
lines this basic idea of partitioning the sample space is the
key. Instead of retreating to the abstract (Kolmogorov measures on 
subsets) machinery of
Table \ref{kol1}, we find a different way ahead by carefully selecting
the intervals $I^{(j)}$ which partition the sample space
$\Omega=\R^B$.

\section{Riemann Sums}

An idea of what is involved in this can be obtained by recalling
the role of Riemann sums in basic integration theory. Suppose for
simplicity that the sample space $\Omega$ is the interval $[a,b[
\subset \R$  and the random variable $f(x)$ is given by $f:\Omega
\rightarrow \R$; and suppose $F: \mathcal{I} \rightarrow [0,1]$
where $\mathcal{I}$ is the family of subintervals $I\subseteq
\Omega = [a,b[$.

We can interpret $F$ as the probability distribution function of
the underlying random variable $x$, so $F(I)$ is the likelihood
that $x \in I$. As a distribution function, $F$ is finitely
additive on $\mathcal{I}$.

The simplest intuition of likelihood -- as something
intermediate between certainty of non-occurrence and certainty of
occurrence -- implies that likelihoods must be representable as
numbers between 0 and 1. It follows that distribution functions
are finitely additive on $\mathcal{I}$. This immediately lifts the
burden of credulity that (\ref{measure}) imposes on our naive or
``natural'' sense of what probability or likelihood is.

With $f$ a deterministic function of the random variable $x$,  the
random variation of $f(x)$ is our object of investigation. In the
first instance we wish to establish $E(f)$, the expected value of
$f(x)$, as, in some sense, the integral of $f$ with respect to
$F$, which is often estimated as in Table \ref{riem1}.

 Following broadly the scheme of Table \ref{riem1}, we first
 select an arbitrary number $\delta>0$.  Then we choose a finite
number of disjoint intervals $I^1, \ldots , I^n$;
$I^j=[u^{j-1},u^j[$, $a=u^0<u^1< \cdots <u^n=b$, with each
interval $I^j$ satisfying \begin{equation}\label{riem} |I^j|:=
u^j-u^{j-1} < \delta. \end{equation} We then select a
representative $x^j$, $u^{j-1} \leq x^j \leq u^j$,
$1 \leq j \leq n$.

(For simplicity we are using superscript $^j$ instead of
$^{(j)}$ --- for labelling, not exponentiation. The reason for not
using subscript $_j$ is to keep such subscripts available  to
denote dimensions in multi-dimensional variables.)

Then the Riemann (or Riemann-Stieltjes) integral of $f$ with
respect to $F$ exists, with  $\int_a^b f(x)dF = \alpha$, if, given
any $\epsilon >0$, there exists a number $\delta>0$ so that
\begin{equation}\label{int1}
\left| \sum_{j=1}^n f(x^j) F(I^j) - \alpha \right| < \varepsilon
\end{equation}
for every such choice of $x^j$, $I^j$ satisfying (\ref{riem}),
$1\leq j \leq n$.

If we could succeed in creating a theory of random variation along
these lines then we could reasonably declare that the expectation
$E^F(f)$ of the random variable $f(x)$, relative to the
distribution function $F(I)$, is $\int_a^b f(x)dF$ whenever the
latter exists in the sense of (\ref{int1}). (In fact this
statement is true, but a justification of it takes us deep into
the Kolmogorov theory of probability and random variation. A
different justification is given in this article.)

But (\ref{riem}) and (\ref{int1}) on their own do not yield an
adequate theory of random variation. For one thing, it is well
known that not every Lebesgue integrable function is Riemann
integrable. So in this sense at least, Table \ref{kol1} goes further
than Table \ref{riem1} and relation (\ref{int1}).

More importantly, any theory of random variation must contain
results such as Central Limit Theorems and Laws of Large Numbers,
which are the core of our understanding of random variation, and
the proofs of such results require theorems like the Dominated
Convergence Theorem, which are available for Table \ref{kol1} and
Lebesgue integrals, but which are not available for the ordinary
Riemann integrals of Table \ref{riem1} and (\ref{int1}).

However, before we take further steps towards the generalization
of the Riemann integral (\ref{int1}) which will give us what we
need, let us pause to give further consideration to data
encoding.

Though the classes $I^j$ used in (\ref{int1}) above are not
required to be of equal length, it is certainly consistent with
(\ref{int1}) to partition the sample data into equal classes. To
see this, choose $n$ so that $ (b-a)/n < \delta$, and then choose
each $u^j$ so that $u^j - u^{j-1} = (b-a)/n$. Then $I^j =
[u^{j-1}, u^j[$ ($1\leq j \leq n$) gives us a partition of $\Omega
= [a,b[$ in which each $I^j$ has the same length $ (b-a)/n$.

We could also, in principle, obtain quantile classification of the
data by this method of $\delta$-partitioning. Suppose we want
decile classification; that is, $[a,b[ = I^1 \cup \cdots \cup I^n$
with $F(I^j) = 0.1$, $1 \leq j \leq n$. This is possible, since
the function $F(u) := F([a,u[)$ is monotone increasing and
continuous for almost all $u \in ]a,b[$, and hence there exist
$u^j$ such that $F(u^j) = j/10$ for $1 \leq j \leq 10$. So if
$\delta$ happens to be greater than  $\max \{u^j - u^{j-1}: 1 \leq
j \leq 10\}$, then the decile classification satisfies $|I^j| =
u^j - u^{j-1} < \delta$ for $1 \leq j \leq 10$. (This argument
merely establishes the existence of such a classification.
Actually determining quantile points for a particular distribution
function requires \emph{ad hoc} consideration of the distribution
function in question.)

In fact, this focus on the system of data encoding is the
avenue to a rigorous theory of random variation  within a Riemann
framework, as we shall now see.

\section{The Generalized Riemann Integral}

In the previous section we took the sample space $\Omega$ to be
$[a,b[$. As our attention from here on is going to be (below in the 
application study) increasingly
focussed on counts or frequencies, which are
non-negative, we will take the sample space to be $\R_+ = ]0,
\infty[$, or a multiple Cartesian product of $\R_+$ by itself.

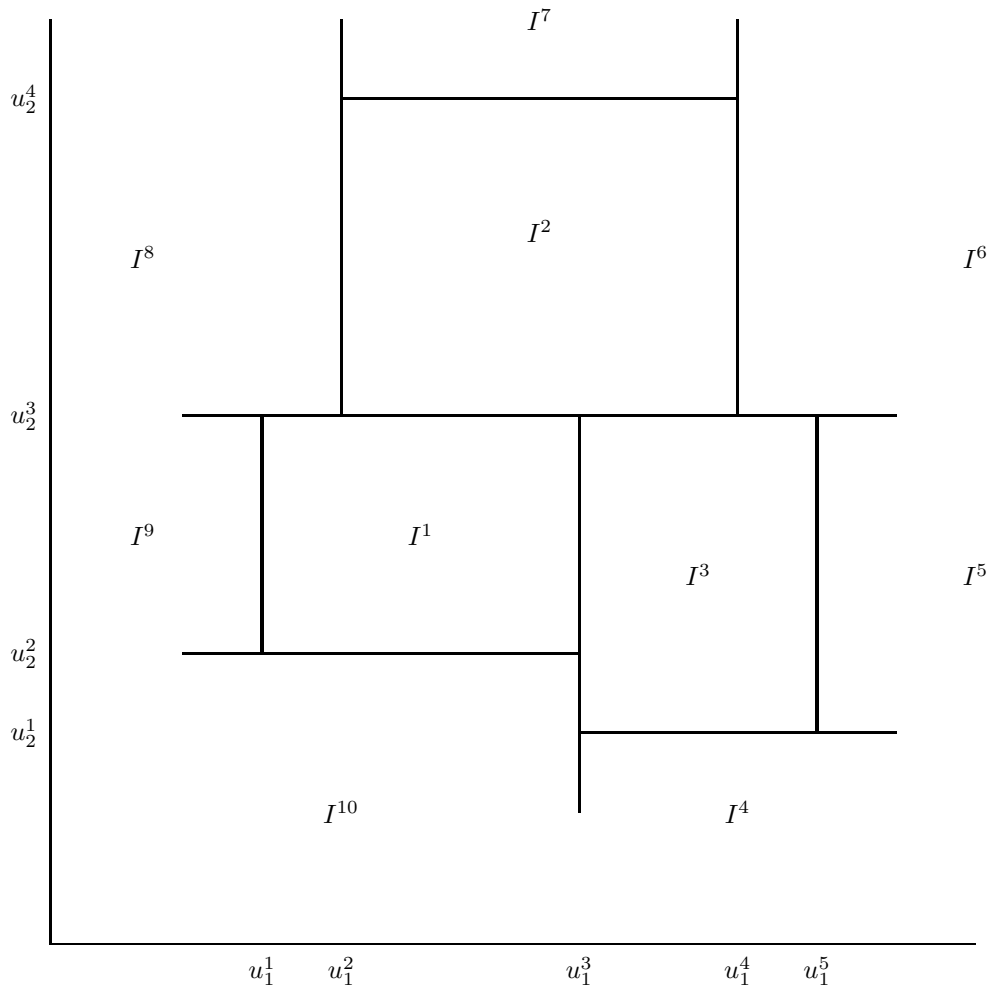
\begin{figure}
\thicklines

\begin{picture}(400,420)

{\thinlines \put(10,10){\line(1,0){350}}
\put(10,10){\line(0,1){350}} }

\put(60,120){\line(1,0){150}} \put(60,210){\line(1,0){270}}
\put(120,330){\line(1,0){150}} \put(210,90){\line(1,0){120}}
\put(90,120){\line(0,1){90}} \put(120,210){\line(0,1){150}}
\put(210,60){\line(0,1){150}} \put(270,210){\line(0,1){150}}
\put(300,90){\line(0,1){120}}


\put(90,0){\makebox(0,0){$u_1^1$}}
\put(120,0){\makebox(0,0){$u_1^2$}}
\put(210,0){\makebox(0,0){$u_1^3$}}
\put(270,0){\makebox(0,0){$u_1^4$}}
\put(300,0){\makebox(0,0){$u_1^5$}}

\put(0,90){\makebox(0,0){$u^1_2$}}
\put(0,120){\makebox(0,0){$u^2_2$}}
\put(0,210){\makebox(0,0){$u^3_2$}}
\put(0,330){\makebox(0,0){$u^4_2$}}

\put(120,60){\makebox(0,0){$I^{10}$}}
\put(270,60){\makebox(0,0){$I^4$}}
\put(45,165){\makebox(0,0){$I^9$}}
\put(150,165){\makebox(0,0){$I^1$}}
\put(255,150){\makebox(0,0){$I^3$}}
\put(45,270){\makebox(0,0){$I^8$}}
\put(195,280){\makebox(0,0){$I^2$}}
\put(360,270){\makebox(0,0){$I^6$}}
\put(195,360){\makebox(0,0){$I^7$}}
\put(360,150){\makebox(0,0){$I^5$}}

\end{picture}
\caption{Unbounded two-dimensional domain with partition used for 
data encoding.}
\label{pic1}
\end{figure}

Figure \ref{pic1} shows a partition of an unbounded
finite-dimensional domain such as  $\R_+ \times \R_+$.
In this illustration,
\begin{equation}\label{intervals1}\begin{array}{lll}
I^1& =& [u_1^1,u_1^3[ \times [u_2^2,u_2^3[ \\
I^2& =& [u_1^2,u_1^4[ \times [u_2^3,u_2^4[ \\
I^3& =& [u_1^3,u_1^5[ \times [u_2^1,u_2^3[ \\
I^4& =& [u_1^3,\infty[ \times ]0,u_2^1[ \\
I^5& =& [u_1^5,\infty[ \times [u_2^1,u_2^3[ \\
I^6& =& [u_1^4,\infty[ \times [u_2^3,\infty[ \\
I^7& =& [u_1^2,u_1^4[ \times [u_2^4,\infty[ \\
I^8& =& ]0,u_1^2[ \times [u_2^3, \infty[ \\
I^9& =& ]0,u_1^1[ \times [u_2^2,u_2^3[ \\
I^{10}& =& ]0,u_1^3[ \times ]0,u_2^2[.
\end{array}
\end{equation}

For each elementary occurrence $x \in \Omega = \R^n$ ($n$ a
positive integer), let $\delta (x)$ be a positive number. Then an
admissible classification of the sample space, called a
$\delta$\emph{-fine division} of $\Omega$, is a finite collection
\begin{equation}\label{div}
\mathcal{E}_\delta := \{(x^j, I^j)\}_{j=1}^n
\end{equation}
so that $x^j$ is in $I^j$. The $I^j$ are disjoint with union
$\Omega$, and the lengths of the edges (or sides) of each $I^j $
are bounded by $ \delta (x^j)$.

So, referring back to Table \ref{riem1} of elementary
statistics, what we are doing here is selecting the data
classification intervals $I^j$ along with a representative value
$x^j$ from $I^j$.

It is convenient (though not a requirement of the theory) that the
representative value $x^j$ should be a vertex of $I^j$, and that
is how we shall proceed.

In the case of the ordinary Riemann integral in a compact domain
(cf. (\ref{int1})), the positive function $\delta$ is simply a
positive constant, and the bound in question is simply the
condition that each edge of each interval has length less than
$\delta$. Ordinary Riemann integration over unbounded domains, or
domains which contain singularity points of the integrand, is
obtained by means of the \emph{improper Riemann integral} (for
details of which, see Rudin (1970) for instance). In contrast, the
generalized Riemann integral handles all of these situations in
essentially the same way, removing the need for improper
extension. In the illustration in Figure \ref{pic1} above, some of the
edges are infinitely long. The precise sense in which each edge
(finite or infinite) of $I^j$ is bounded by $\delta (x^j)$ is
explained at the end of this section.

The Riemann sum corresponding to (\ref{div}) is
\begin{equation}\label{sum}
(\mathcal{E}_\delta ) \sum f(x) F(I) := \sum_{j=1}^n f(x^j)
F(I^j)
\end{equation}
i.e.\ it is simply the sum over the terms in equation (\ref{div}).
We say that $f$ is \emph{generalized Riemann integrable} with
respect to $F$, with $\int_\Omega f(x)F(I) = \alpha$, if, for each
$\varepsilon >0$, there exists a function $\delta: \Omega
\rightarrow \R_+$ so that, for every $\mathcal{E}_\delta $,
 \begin{equation} \label{int}
\left|  (\mathcal{E}_\delta ) \sum f(x)F(I) -\alpha \right| <
\varepsilon.
\end{equation}
With this step we overcome the two previously mentioned objections
to the use of  Riemann-type integration in a theory of random
variation. Firstly, every function $f$ which is Lebesgue-Stieltjes
integrable in $\Omega$ with respect to $F$ is also generalized
Riemann integrable, in the sense of (\ref{int}). See Gordon (1994)
for a proof of this. Secondly, we have theorems such as the
Dominated Convergence Theorem (see, for example, Gordon, 1994)
which enable us to prove Laws of Large Numbers, Central Limit
Theorems and other results which are needed for a theory of random
variation.

So we can legitimately use the usual language and notation of
probability theory. Thus, the expectation of the random variable
$f(x)$ with respect to the probability distribution function
$F(I)$ is
\[
E^F(f(x)) = \int_\Omega f(x) F(I).
\]
To recapitulate, elementary statistics involves calculations of
the form (\ref{riem1}), often with classes $I$ of equal size or
equal likelihood. We refine this method by carefully selecting the
data classification intervals $I$. In fact our Riemann sum
estimates involve choosing a finite number of occurrences
$\{x^{(1)}, \ldots , x^{(n)}\}$ from $\Omega$ (actually, from the
closure of $\Omega$), and then selecting associated classes
$\{I^{(1)}, \ldots , I^{(n)}\}$, disjoint with union $\Omega$,
with $x^{(j)}$ in $ I^{(j)}$ (or with each $x^{(j)}$ a
vertex of $I^{(j)}$, in the version of the theory that we are
presenting here), such that for each $1 \leq j \leq n$, $I^{(j)}$
is $\delta$-\emph{fine}. The meaning of this is as follows.


Let $\bbR=\R_+\cup \{0, \infty\}$ be $\R_+$ with the points $0$
and $\infty$ adjoined. 
(In the following paragraph,
$x=0$ and $x=\infty$ are given special treatment. Many functions
are undefined for $x=\infty$; and $x=0$ is a singularity for the
function $\ln x$ which may be of use in data normalization -- for 
instance when dealing with astronomy stellar magnitudes or financial ratios.)

Let $I$ be an interval in $\R_+$, of the form
\begin{equation}\label{interval}
]0, v[, \,\,\,\, [u,v[, \,\,\,\, \mbox{or  } [u, \infty[,
\end{equation}
and let $\delta: \bbR \rightarrow ]0,\infty[$ be a positive
function defined for $x \in \bbR$. The function $\delta$ is called
a {\em gauge} in $\R_+$. We say that $I$ is \emph{attached to} $x$
(or \emph{associated with} $x$) if
\begin{equation} \label{tag}
x=0,\,\,\,\, x=u\mbox{ or } v, \,\,\,\,x=\infty
\end{equation}
respectively. If $I$ is attached to $x$ we say that $(x,I)$ is
$\delta$-{\em fine} (or simply that $I$ is $\delta$-fine) if
\begin{equation} \label{fine} v<\delta(x), \,\,\,\,
v-u<\delta(x),\,\,\,\, u> \frac{1}{\delta(x)}
\end{equation}
respectively.

That is what we mean by $\delta$-fineness in one dimension. What
about higher dimensions?

Suppose $I=I_1 \times I_2 \times \cdots \times I_n$ is an interval
of $\R_+^n=\R_+ \times \R_+ \times \cdots \R_+$, each $I_j$ being
a one-dimensional interval of form (\ref{interval}). A point
$x=(x_1, x_2, \ldots , x_n)$ of $\bbR^n$ is attached to $I$ in
$\R_+^n$ if each $x_j$ is attached to $I_j$ in $\R_+$, $1 \leq j
\leq n$. Given a function $\delta: \bbR^n \rightarrow ]0, \infty[$, an
associated pair $(x,I)$ is $\delta$-fine in $\R_+^n$ if each $I_j$
satisfies the relevant condition in (\ref{fine}) with the new
$\delta (x)$. A finite collection of associated $(x,I)$ is a
$\delta$-fine division of $\R_+^n$ if the intervals $I$ are
disjoint with union $\R_+^n$, and if each of the $(x,I)$ is
$\delta$-fine. A proof of the existence of such a $\delta$-fine
division is given in  Henstock (1988), Theorem 4.1.

A glance at Diagram (\ref{pic1}) above will show that many of
points $x$ involved in a division of $\R_+^n$ (vertices of the
partitioning intervals), which correspond to the representative
occurrences $x^{(j)}$ of the data encoding  in Table \ref{riem1},
will belong to $\bbR^n \setminus \R_+^n$; in other words $x$ may
have some components $x_j$ equal to $0$ or $\infty$. The special
arrangements we have made for such points, in (\ref{fine}) above,
are in anticipation of the singularities that are present at such
points in the expressions that arise in our data encoding 
problem. These arrangements, which are characteristic of
generalized Riemann integration, forestall any need for the kind
of improper extensions which are needed in other integration
theories.

\section{But Where Is The Calculus of Probabilities?}

There are certain familiar landmarks in the study of probability
theory and its offshoots such as the calculus of probabilities,
which has not entered into the discussion thus far. The key point
in this calculus is the relationship
\[
P(\cup_{j=1}^\infty A_j) =\sum_{j=1}^\infty P(A_j).
\]
In fact the set-functions $P$ and their calculus are not used as
the basis of the generalized Riemann approach to the study of
random variation. Instead, the basis is the simpler set-functions
$F$, defined only on intervals, and finitely additive on them.

But, as mentioned earlier, an outcome of the generalized Riemann
approach is that we can recover set-functions defined on sets
(including the measurable sets of the Kolmogorov theory) which are
more general than intervals, and we can recover the probability
calculus which is associated with them.

To see this, suppose $A \subseteq \Omega$ is such that
$\int_\Omega \mathbf{1}_A(x) F(I)$ exists in the sense of
(\ref{int}). Then define
\begin{equation}\label{prob}
P_F(A)=\int_\Omega \mathbf{1}_A(x) F(I),
\end{equation}
and we can easily deduce from the Dominated Convergence Theorem
for generalized Riemann integrals, that for disjoint $A_j$ for
which $P_F(A_j)$ exists,
\[
P_F(\cup_{j=1}^\infty A_j) =\sum_{j=1}^\infty P_F(A_j).
\]
Other familiar properties of the calculus of probabilities are
easily deduced from (\ref{prob}).

Since every Lebesgue integrable function is also generalized
Riemann integrable (Gordon, 1994), every result obtained by
Lebesgue integration is also valid for generalized Riemann
integration. So in this sense, the generalized Riemann theory of
random variation is an extension or generalization of the theory
developed by Kolmogorov, Levy, It\^{o} and others.

However the kind of argument which is natural for Lebesgue
integration is different from that which would naturally be used
in generalized Riemann integration, so it is more productive in
the latter case to develop the theory of random variation from
first principles on Riemann lines. Some pointers to such a
development are given in (Muldowney, 1999).

Many of the standard distributions (normal,  exponential and
others)  are mathematically elementary, and the expected or
average values of random variables, with respect to these
distributions---whether computed by means of the generalized
Riemann or Lebesgue methods---often reduce to Riemann or
Riemann-Stieltjes integrals. Many aspects of these distributions
can be discovered with ordinary Riemann integration. But it is
their existence as generalized Riemann integrals, possessing
properties such as the Dominated Convergence Theorem and Fubini's
Theorem, that gives us access to a full-blown theory of random
variation.

\section{Marginal Distributions and Statistical Independence}
When random variables $\{x_t\}_{t \in B}$ are being considered
jointly, their \emph{marginal} behavior is a primary
consideration. This means examining the joint behavior of any
finite subset of the variables, the remaining ones (whether
finitely or infinitely many) being arbitrary or left out of
consideration. Thus we are led to families
\[
\{x_t: t \in N\} _{N\subseteq B}
\]
where the sets $N$ belong to the family $\mathcal{F}$ of finite
subsets of $B$, the set $B$ being itself finite or infinite. (When
$B$ is infinite the family $(x_t)_{t \in B}$ is often called a
\emph{process} or \emph{stochastic process}, especially when the
variable $t$ represents time. We will write the random variable
$x_t$ as $x(t)$ depending on the context; likewise $x_{t_j} =
x(t_j) = x_j$.)  In the following discussion we will suppose, for 
illustrative purposes, that for each $t$ the domain of values of 
$x_t$ is the set $\R_+$ of positive numbers.  This would apply if, 
for instance, $(x_t)$ is price history, $t \in B$.

The marginal behavior of a process is specified by marginal distribution
functions.  
The marginal distribution function of the
random variable or process $x_B = (x_t)_{t \in B}$, for any finite
subset $N = \{t_1, t_2, \ldots , t_n \} \subseteq B$, is the
function
\begin{equation} \label{dist1}
F_{(x_1, x_2, \ldots , x_n)} (I_1 \times I_2 \times \cdots \times
I_n)
\end{equation}
defined on the intervals $I_1 \times \cdots \times I_n$ of
$\R_+^n$, which we interpret as the likelihood that the random
variable $x_j $ takes a value in the one-dimensional interval
$I_j$ for each $j$, $1 \leq j \leq n$; with the remaining random
variables $x_t$ arbitrary for $t \in B \setminus N$.

One of the uses to which the marginal behavior is put is to
determine the presence or absence of \emph{independence}. The
family of random variables $(x_t)_{t \in B}$ is \emph{independent}
if the marginal distribution functions satisfy
\[
F_{(x_1, x_2, \ldots , x_n)} (I_1 \times I_2 \times \cdots \times
I_n) = F_{x_1}(I_1) \times F_{x_2}(I_2) \times \cdots F_{x_n}(I_n)
\]
for every finite subset $N= \{t_1, \ldots , t_n\} \subseteq B$.
That is, the likelihood that the random variables $x_{t_1}$,
$x_{t_2}$, $\ldots $, $x_{t_n}$ jointly take values in $I_1$,
$I_2$ $ \ldots$, $I_n$ (with $x_t$ arbitrary for $t \in B
\setminus N$) is the product over $j=1,2, \ldots ,n$ of the
likelihoods of $x_{t_j}$ belonging to $I_j$ (with $x_t$ arbitrary
for $t \neq t_j$, $j=1,2, \ldots , n$) for every choice of such
intervals, and for every choice of finite subset $N$ of $B$.

Of course, if $B$ is itself finite, it is sufficient to consider
only $N=B$ in order to establish whether or not the random
variables are independent.

\section{Cylindrical Intervals to Support Infinite Dimensional Spaces}
\label{cylint}

When $B$ is  infinite (so the random variable $x=(x(t))_{t\in B}$
is a stochastic process), it is usual to define the distribution
of $x$ as the family of distribution functions
\begin{equation}\label{cyl1}
\left\{F_{(x(t_1), x(t_2), \ldots , x(t_n))}(I_1 \times I_2 \times
\cdots \times I_n): \{t_1, t_2, \ldots , t_n \} \subset B \right\}
\end{equation}
This is somewhat awkward, since up to this point the distribution
of a random variable has been given as a single function defined
on intervals of the sample space, and not as a family of
functions. However we can tidy up this awkwardness as follows.

Firstly, the sample space $\Omega$ is now the Cartesian product
$\prod_B \R_+ = \R_+^B$. Let $\mathcal{F}$ denote the family of
finite subsets $N= \{t_1, t_2, \ldots , t_n\}$ of $B$. Then for
any $N \in \mathcal{F}$, the set
\[
I[N] := I_{t_1} \times I_{t_2}\times \cdots \times I_{t_n} \times
\prod\{\R_+: B \setminus N \}
\]
is called a \emph{cylindrical interval}. Taking all choices of $N
\in \mathcal{F}$ and all choices of one-dimensional intervals
$I_j$ ($t_j \in N$), denote the resulting class of cylindrical
intervals by $\mathcal{I}$. These cylindrical intervals are the
subsets of the sample space that we need to define the
distribution function $F$ of $x$ in $\R_+^B$:
\begin{equation}\label{cylfn}
F(I[N]) := F_{(x(t_1), x(t_2), \ldots , x(t_n))}(I_{t_1} \times
I_{t_2} \times \cdots \times I_{t_n})
\end{equation}
for every $N \in \mathcal{F}$ and every $I[N] \in \mathcal{I}$.

By thus defining the distribution function $F$ (of the underlying
random variable $x \in \R_+^B$) on the family of subsets
$\mathcal{I}$ (the cylindrical intervals) of $\R_+^B$, we are in
conformity with the system used for describing distribution
functions in finite-dimensional sample spaces.

As in the elementary situation of Table \ref{riem1}, it naturally
follows, if we want to estimate the expected value of some
deterministic function of the random variable (or process)
$(x(t))_{t \in B}$, that the joint sample space $\Omega = \R_+^B$ of
the individual random variables $x(t)$ should be partitioned by
means of cylindrical intervals $I[N]$.

To demonstrate such a partition, we suppose $B$ is the time
interval $]\tau,T]$, so the sample space $\Omega$ is $\R_+^B =
\prod_{t \in ]\tau,T]} \R_+ = \R_+^{]\tau,T]}$. Suppose \[\tau=t_0
< t_1<t_2 < \cdots < t_n=T,\] and, with $N$ denoting $\{t_1, t_2,
\ldots , t_n\}$, suppose \[I[N] = I_1 \times I_2 \times \cdots
\times I_n \times \R_+^{B \setminus N}\] is one of the cylindrical
intervals forming a partition of $\R_+^B$.

\begin{figure}
\begin{picture}(400,420)

\put(60,120){\line(1,0){150}} \put(60,210){\line(1,0){270}}
\put(120,330){\line(1,0){150}} \put(210,90){\line(1,0){120}}
\put(60,240){\line(1,0){270}}

\put(90,120){\line(0,1){90}} \put(120,240){\line(0,1){120}}
\put(210,60){\line(0,1){150}} \put(270,240){\line(0,1){120}}
\put(300,90){\line(0,1){120}}


\put(90,0){\makebox(0,0){$u_1^1$}}
\put(120,0){\makebox(0,0){$u_1^2$}}
\put(210,0){\makebox(0,0){$u_1^3$}}
\put(270,0){\makebox(0,0){$u_1^4$}}
\put(300,0){\makebox(0,0){$u_1^5$}}

\put(0,90){\makebox(0,0){$u^1_2$}}
\put(0,120){\makebox(0,0){$u^2_2$}}
\put(0,210){\makebox(0,0){$u^3_2$}}
\put(0,240){\makebox(0,0){$u^4_2$}}
\put(0,330){\makebox(0,0){$u^5_2$}}

\put(150,60){\makebox(0,0){$I^{10}$}}
\put(240,60){\makebox(0,0){$I^4$}}
\put(45,165){\makebox(0,0){$I^9$}}
\put(150,165){\makebox(0,0){$I^1$}}
\put(255,150){\makebox(0,0){$I^3$}}
\put(45,270){\makebox(0,0){$I^8$}}
\put(195,280){\makebox(0,0){$I^2$}}
\put(195,225){\makebox(0,0){$I^{11}$}}
\put(360,270){\makebox(0,0){$I^6$}}
\put(195,360){\makebox(0,0){$I^7$}}
\put(360,150){\makebox(0,0){$I^5$}}

\put(60,105){\line(2,1){40}}
 \put(60,195){\line(2,1){40}}
\put(90,225){\line(2,1){40}} \put(240,315){\line(2,1){40}}
\put(90,315){\line(2,1){40}} \put(240,225){\line(2,1){40}}
\put(180,195){\line(2,1){40}} \put(270,195){\line(2,1){40}}
\put(180,105){\line(2,1){40}} \put(180,75){\line(2,1){40}}
\put(270,75){\line(2,1){40}}

\end{picture}
\caption{As for Figure \ref{pic1}, unbounded two dimensional domain with 
partition used for data encoding, illustrating the use of different 
restricted dimensions.}
\label{pic2}
\end{figure}
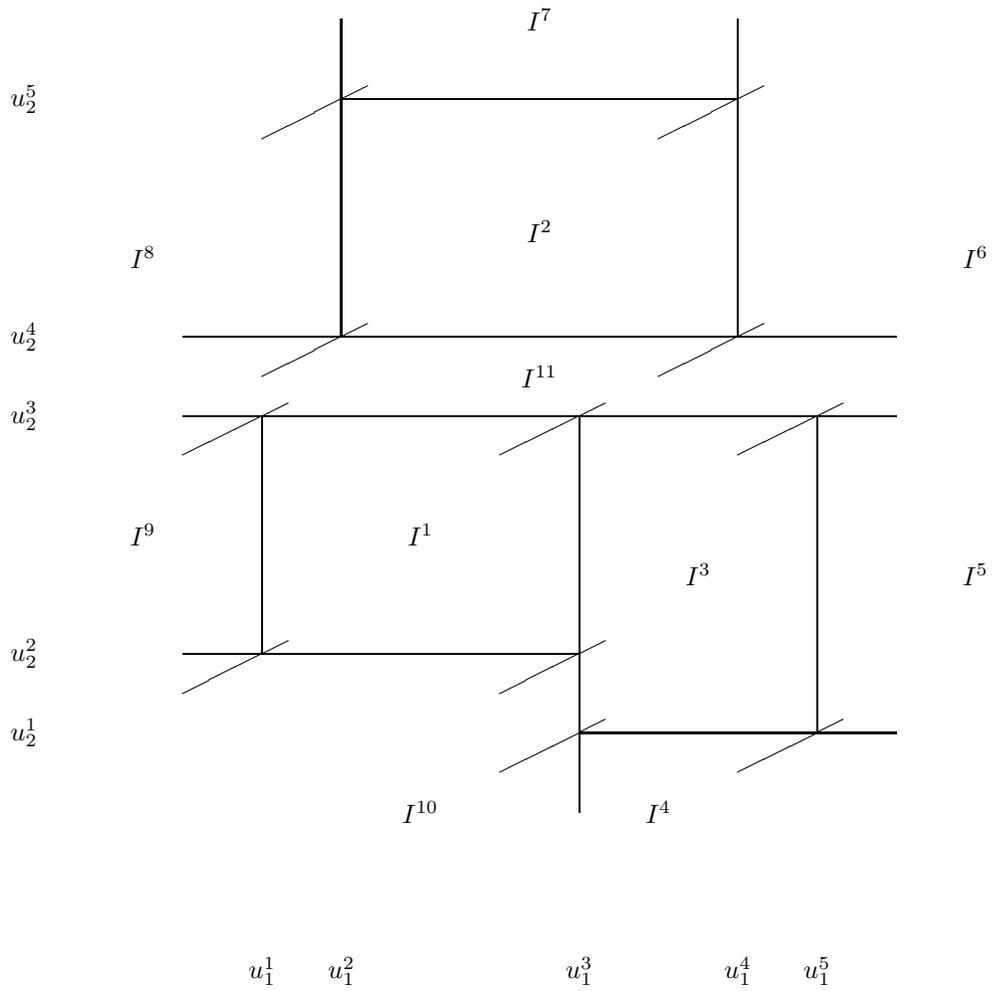

In Figure \ref{pic2}, we can show only three
dimensions. As in Figure \ref{pic1}, the fact that the sample space is
unbounded in each of its separate dimensions means that many of
the partitioning intervals have associated points with one or more
components equal to $0$ or $\infty$. 
We have terms $\ln x_j$ in the integrand which are
undefined for $x_j=0$, just as $\ln \infty$ is undefined. In
generalized Riemann integration, any intervals involving a
singularity must have the point of singularity as the attached or
associated point. By arranging things in this way, generalized
Riemann integration avoids having to resort to the improper or
Cauchy extensions when the integrand involves a point of
singularity.

 In contrast to Figure \ref{pic1}, 
the partitioning intervals may have different
restricted dimensions. For instance, in Figure \ref{pic2}, the
cylindrical interval $I^{11}$ is restricted only in the vertical
direction $t_2$; and is unrestricted in the horizontal direction
$t_1$ and in each of the infinitely many other directions $t \in B
\setminus \{ t_1, t_2\}$ (of which only one of the directions
perpendicular to both $t_1$ and $t_2$ is shown in the diagram).
This is a particular feature of partitioning infinite-dimensional
domains by means of infinite-dimensional cylindrical intervals,
which we must take account of when we construct Riemann sums of
integrands over such partitions.

In this illustration (Figure \ref{pic2})
the cylindrical intervals mostly correspond
to the finite-dimensional intervals of (\ref{intervals1}), but an
extra one, $I^{11}$, has been included to demonstrate that the
restricted dimensions of the cylindrical intervals do not all have
to be the same in a partition of an infinite-dimensional space.
(Of course this is also true for finite dimensional spaces. We
could have included an interval corresponding to $I^{11}$ in
(\ref{intervals1}), but in partitioning for Riemann sum estimates
in the finite-dimensional case, these kind of intervals can be
avoided and nothing is gained by admitting them. But in
partitioning infinite-dimensional spaces they cannot be avoided.)

The intervals in Figure \ref{pic2} are:
\begin{equation}\label{intervals2}
\begin{array}{lll}
I^1& =& [u_1^1,u_1^3[ \times [u_2^2,u_2^3[ \times \prod \{\R_+: t \in B \setminus \{t_1,t_2\} \},\\
I^2& =& [u_1^2,u_1^4[ \times [u_2^4,u_2^5[  \times \prod \{\R_+: t \in B \setminus \{t_1,t_2\} \},\\
I^3& =& [u_1^3,u_1^5[ \times [u_2^1,u_2^3[  \times \prod \{\R_+: t \in B\setminus \{t_1,t_2\} \},\\
I^4& =& [u_1^3,\infty[ \times ]0,u_2^1[  \times \prod \{\R_+: t \in B\setminus \{t_1,t_2\} \},\\
I^5& =& [u_1^5,\infty[ \times [u_2^1,u_2^3[  \times \{\prod \R_+: t \in B\setminus \{t_1,t_2\} \},\\
I^6& =& [u_1^4,\infty[ \times [u_2^4,\infty[  \times \{\prod \R_+: t \in B\setminus \{t_1,t_2\} \},\\
I^7& =& [u_1^2,u_1^4[ \times [u_2^5,\infty[  \times \{\prod \R_+: t \in B\setminus \{t_1,t_2\} \},\\
I^8& =& ]0,u_1^2[ \times [u_2^4, \infty[  \times \{\prod \R_+: t \in B\setminus \{t_1,t_2\} \},\\
I^9& =& ]0,u_1^1[ \times [u_2^2,u_2^3[  \times \{\prod \R_+: t \in B\setminus \{t_1,t_2\} \},\\
I^{10}& =& ]0,u_1^3[ \times ]0,u_2^2[ \times \{\prod \R_+ : t \in B\setminus \{t_1,t_2\} \},\\
I^{11}& =&]u_2^3,u_2^4[ \times \prod \{\R_+: t \in B, t \neq
t_2\}.
\end{array}
 \end{equation}

Criteria (\ref{int}),
 (\ref{GenRiemInt2}) place no a priori conditions on the
functions $f$ and $F$ in the integrand when we test it for
integrability. There are no required or preferred kinds of
function. It is true that we have required $F$ to be finitely
additive, but this is related to our secondary purpose of
constructing an alternative to the Kolmogorov theory of
probability and random variation. Of course, in meeting the
criteria (\ref{int}), (\ref{GenRiemInt2}), any good properties
possessed by $f$ and $F$ may come into play in order to give us
a good encoding.  The foregoing remarks may be translated into 
language that is more appropriate for statistical data analysis: 
there is no necessary a priori morphology for the data cloud to be analyzed; or
there is no necessary a priori model or distribution for the data.  

\section{A Theory of Joint Variation of Infinitely Many Random Variables}
\label{infint}

 As discussed earlier, the Riemann sum approach can
be adapted so that it yields a theory of random variation which
meets the theoretical and practical needs of analysis.

The adaptation that is needed when only a finite number of random
variables is involved has been explained already.

But how can it be adapted to the situation when there are
infinitely many random variables to be considered jointly?
What kind of Riemann sums are appropriate in a
rigorous theory of joint variation of infinitely many variables?

In other words, what kind of partitions are permitted in forming
the Riemann sum approximation to the expected value of a random
variable which depends on infinitely many underlying random
variables?

In ordinary Riemann integration we form Riemann sums by choosing
partitions whose finite-dimensional intervals have edges (sides)
which are bounded by a positive constant $\delta$. Then we make
$\delta$ successively smaller. Likewise for generalized Riemann
integration, where the constant $\delta$ is replaced by a positive
function $\delta(x)$. In any case, we are choosing successive
partitions in which the component intervals successively
``shrink'' in some sense.

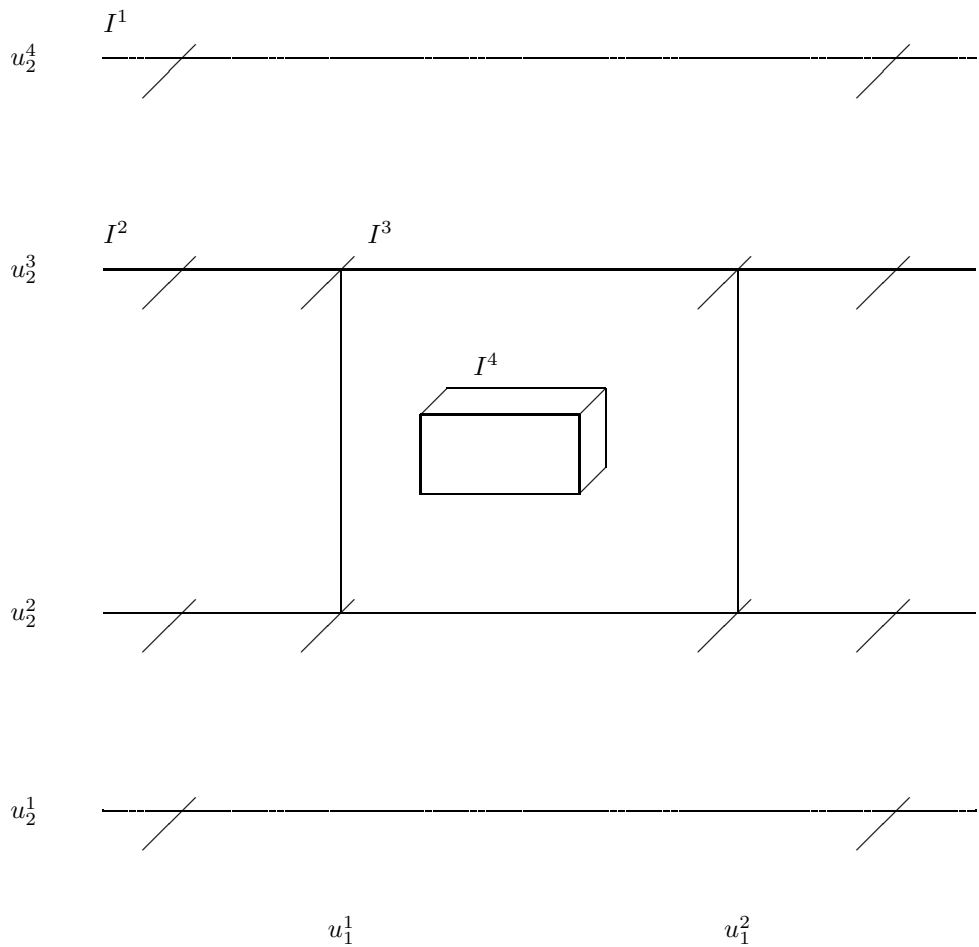
\begin{figure}
\begin{picture}(400,420)

\put(30,45){\dashbox{.5}(330,0)} \put(30,330){\dashbox{.5}(330,0)}

\put(45,30){\line(1,1){20} } \put(45,315){\line(1,1){20} }
\put(315,30){\line(1,1){20} } \put(315,315){\line(1,1){20} }

\put(0,45){\makebox(0,0){$u_2^1$}}
\put(0,330){\makebox(0,0){$u_2^4$}}

\put(30,120){\line(1,0){330}   } \put(30,250){\line(1,0){330}   }

\put(0,120){\makebox(0,0){$u_2^2$}}
\put(0,250){\makebox(0,0){$u_2^3$}}
 \put(45,105){\line(1,1){20} }
\put(45,235){\line(1,1){20} } \put(315,105){\line(1,1){20} }
\put(315,235){\line(1,1){20} }

\put(120,120){\line(1,0){150}} \put(120,250){\line(1,0){150}}
\put(120,120){\line(0,1){130}} \put(270,120){\line(0,1){130}}

\put(120,0){\makebox(0,0){$u^1_1$}}
\put(270,0){\makebox(0,0){$u^2_1$}}

\put(105,105){\line(1,1){20} } \put(105,235){\line(1,1){20} }
\put(255,105){\line(1,1){20} } \put(255,235){\line(1,1){20} }

\put(150,165){\line(1,0){60}} \put(150,195){\line(1,0){60}}
\put(150,165){\line(0,1){30}} \put(210,165){\line(0,1){30}}

\put(210,195){\line(1,1){10}} \put(210,165){\line(1,1){10} }

\put(160,205){\line(1,0){60}} \put(220,175){\line(0,1){30}}
\put(150,195){\line(1,1){10} }

\put(30,340){\makebox(0,0)[bl]{$I^1$}}
\put(30,260){\makebox(0,0)[bl]{$I^2$}}
\put(130,260){\makebox(0,0)[bl]{$I^3$}}
\put(170,210){\makebox(0,0)[bl]{$I^4$}}

\end{picture}
\caption{Illustration of different ways in which a cylandrical interval
can be a subset of a larger cylandrical interval; and hence how data 
encoding level resolution is supported.}
\label{pic4}
\end{figure}

For the infinite-dimensional situation, we seek likewise to
``shrink'' the cylindrical intervals $I[N]$ of which successive
partitions are composed. In Figure \ref{pic4} we show
different ways in which a cylindrical interval can be a subset of
a larger cylindrical interval, and hence seek to establish
effective rules by which intervals of successive partitions can be
made successively smaller.

Let the horizontal direction in Figure \ref{pic4} be denoted $t_1$, denote
the vertical direction  by $t_2$, and denote the
direction perpendicular to both by $t_3$. Let $B$ denote the set
of all the dimensions, or mutually perpendicular directions, of
the domain $\R_+^B$. Then $I^1$ is $[u_2^1,u_2^4[ \times \prod\{
\R_+: {t \in B,\; t \neq t_2}\}$. The interval $I^2= [u_2^2,u_2^3[
\times \prod \{ \R_+:{t \in B,\; t \neq t_2}\}$ is a subinterval
of $I^1$, in which the side corresponding to restricted dimension
$t_2$ is shorter than the corresponding side of $I^1$. This kind
of ``shrinking'' is familiar from finite-dimensional Riemann
integration. We get it by imposing a condition that the sides of
the intervals be less than some positive function $\delta$, and
then taking $\delta$ successively smaller.

Now consider $I^3 = [u_1^1, u_1^2[ \times [u_2^2,u_2^3[ \times
\prod \{\R_+:t \in B \setminus \{t_1,t_2\}\}$, which is a subset
of $I^2$, in which the length of the restricted sides is the same
as the length of the restricted side of $I^2$; but in which there
is an additional restricted dimension $t_1$. Here we obtain
shrinking, without changing $\delta$, but by requiring the
interval to have additional restricted dimensions. We can do this
by specifying some minimal finite set of dimensions in which the
interval must be restricted. (We may allow the interval to be
restricted in additional dimensions outside of this minimal set;
just as the sides can be as small as we like provided their length
is bounded by $\delta$.) Then we can obtain shrinking of the
intervals by increasing without limit the number of elements in
this minimal finite set, just as we can obtain shrinking by
decreasing towards zero the size of the $\delta$ which bounds the
lengths of the restricted sides.

If we compare $I^4$ with $I^2$ we see both factors at work
simultaneously -- increased restricted dimensions and reduced
length of sides.

This provides us with the intuition we need to construct
appropriate rules for forming partitions for Riemann sums in
infinite-dimensional spaces.

As before, suppose $B$ is a set with a possibly infinite number of
elements. Let $\mathcal{F}$ denote the family of finite subsets
$N$ of $B$. Let a typical  $N \in \mathcal{F}$ be denoted $\{t_1,
t_2, \ldots ,t_n\}$. Suppose the sample space is $\Omega =
\R_+^B$. For $N \in \mathcal{F}$, let $\R_+^N$ denote the
projection of $\Omega$ into the finite set $N$. Suppose $I_j$ is
an interval of type (\ref{interval}) in $\R_+^{\{t_j\}}$. Then
$I_1 \times I_2 \times \cdots \times I_n \times \R_+^{B\setminus
N}$ is a cylindrical interval, denoted $I[N]$. As before, let
$\mathcal{I}$ denote the class of cylindrical intervals obtained
through all choices of $N \in \mathcal{F}$, and all choices of
intervals $I_j$ of type (\ref{interval}), for each $t_j \in N$. A
point $x \in \bbR^B$ is associated with a cylindrical interval
$I[N]$ if, for each $t_j \in N$, the component $x_j = x(t_j)$ is
associated with $I_j$ in the sense of (\ref{tag}). A finite
collection $\mathcal{E}$ of associated pairs $(x,I[N])$ is a
\emph{division} of $\R_+^B$ if the finite number of the
cylindrical intervals $I[N]$ form a partition of $\R_+^B$; that
is, if they are disjoint with union $\R_+^B$.

Now define functions $\delta_N$ and $L$  as follows. Let $L:
\bbR^B \mapsto \mathcal{F}$, and for each $N \in \mathcal{F}$ let
$\delta_N: \bbR^N \mapsto ]0, \infty[$. The mapping $L$ is defined
on the set of associated points of the cylindrical intervals $I[N]
\in \mathcal{I}$; and, for each $N \in \mathcal{F}$, the mapping
$\delta_N$ is a function defined on the set of associated points
of intervals $I_1 \times \cdots \times I_n$ in $\R_+^N$.

The sets $L(x)$ and the numbers $\delta_N(x_1, \ldots ,x_n)$
determine the kinds of cylindrical intervals, partitioning the
sample space,  which we permit in forming Riemann sums.

A set $L(x) \in \mathcal{F}$ determines a minimal set of
restricted dimensions which must be possessed by any cylindrical
interval $I[N]$ associated with $x$. In other words, we require
that $N \supseteq L(x)$. The numbers $\delta_N (x_1, \ldots ,x_n)$
form the bounds on the lengths of the restricted faces of the
cylindrical intervals $I[N]$ associated with $x$. Formally, the
role of $L$ and $\delta_N$ is as follows.

For any choice of $L$ and any choice of the family
$\{\delta_N\}_{N \in \mathcal{F}}$, let $\gamma$ denote
$(L,\{\delta_N\}_{N \in \mathcal{F}})$. We call $\gamma$ a
\emph{gauge} in $\R_+^B$. The class of all gauges is obtained by
varying the choices of the mappings $L$ and $\delta_N$.

Given a gauge $\gamma$, an associated pair $(x,I[N])$ is
$\gamma$-\emph{fine} provided $ N\supseteq L(x)$, and provided,
for each $t_j \in N$, $ (x_j,I_j)$ is $\delta_N$-fine, satisfying
the relevant condition in (\ref{fine}) with $\delta_N(x_1, x_2,
\ldots x_n)$ in place of $\delta(x)$.

Given a random variable, or function $f$ of $x$, with a
probability distribution function $F$ defined on the cylindrical
intervals $I[N]$ of $\mathcal{I}$, the integrand $f(x)F(I[N])$ is
integrable in $\R_+^B$, with   $\int_{\R_+^B} f(x)F(I[N]) =
\alpha$, if, given $\varepsilon>0$, there exists a gauge $\gamma$
so that, for every $\gamma$-fine division $\mathcal{E}_\gamma$ of
$\R_+^B$, the corresponding Riemann sum satisfies
\begin{equation}\label{GenRiemInt2}
\left|(\mathcal{E}_\gamma) \sum f(x)F(I[N]) - \alpha \right| <
\varepsilon.
\end{equation}
If $B$ is finite, this definition reduces to definition
(\ref{int}), because, as each $L(x)$ increases, in this case it is
not ``without limit''; as eventually $L(x)=B$ for all $x$, and
then (\ref{GenRiemInt2}) is equivalent to (\ref{int}). Also
(\ref{GenRiemInt2}) yields results such as Fubini's Theorem and
the Dominated Convergence Theorem (see Muldowney, 1988) which are
needed for the theory of joint variation of infinitely many random
variables.

\section{Application to Financial Data Analysis}

In a number of papers, Muldowney (2000/2001, 2002, 2005) has explored 
expectation and, more generally, integral properties of the Black-Scholes
model of derivative asset pricing.  In the application studied in this 
article, we will consider the finding of structure in empirical financial 
data.  For this we will use correspondence analysis, because it provides 
an integrated tool set for assessing departure from standard behavior in the 
data.  

Correspondence analysis is a data analysis approach based on low-dimensional
spatial projection.  Unlike other such approaches, it particularly well 
caters for qualitative or categorical input data, including counts.  
Hence it is an ideal example of our view that generalized Riemann integration
offers a solid theoretical framework on which to base such an analysis.

Our objectives in this analysis are to take data recoding as proposed
in Ross (2003) and study it as a type of coding commonly used in 
correspondence analysis.  Ross (2003) uses input data recoding to find 
faint patterns in otherwise apparently structureless data.  The implications
of doing this are important: we wish to know if such data recoding can be
applied in general to apparently structureless financial or other data 
streams.  

More particularly our objectives are to assess the following:

\begin{enumerate}
\item Using categorical or qualitative coding may allow structure, 
imperceptible with quantitative data, to be discovered.
Quantile-based categorical coding (i.e., the uniform prior case) 
 has beneficial properties, as will be demonstrated.  
But the issue of  appropriate coding granularity, or 
scale of problem representation, remains, and we will address this issue 
below.
\item In the case of a time-varying data signal (which also holds for 
spatial data, {\em mutatis mutandis}) non-respect of stationarity should be
checked for: the consistency of our results will inform us about 
stationarity present in our data.
More generally, structures 
(or models or associations or relationships) found in 
our data are validated through consistency of results obtained using 
subsets of the population studied.
\item Departure from average behavior is made easy in the analysis 
framework adopted.  This amounts to fingerprinting the data, i.e.\ 
determining patterns in the data that are characteristic of it. 
\end{enumerate}

\section{Searching for Structure in Price Processes}

\subsection{Data Transformation and Coding}

Using crude oil data, Ross (2003) shows how structure can 
be found in apparently geometric Brownian motion, through
data recoding.  Considering monthly oil price values, $P(i)$, 
and then $L(i) = \log (P(i))$, and finally $D(i) = L(i) - 
L(i-1)$, a histogram of $D(i)$ for all $i$ should approximate
a Gaussian.  The following recoding, though, gives rise to 
a somewhat different picture: response categories or states 
1, 2, 3, 4 are used for values of $D(i)$ less than or equal to 
$-0.01$, between the latter and 0, from 0 to $0.01$, and 
greater than the latter.  Then a cross-tabulation of states
1 through 4 for $y_{t+1}$, against states 1 through 4 for
$y_t$, is determined.  The cross-tabulation can be expressed
as a percentage.  Under geometric Brownian motion, one 
would expect constant percentages.  This is not what is found.
Instead there is appreciable structure in the contingency 
table.  

Ross (2003) pursues exploration of a geometric Brownian motion
justification for Black-Scholes option cost.   States-based pricing
leads to greater precision compared to a one-state alternative.
The number of states is left open with both a 4-state and a 
6-state analysis discussed (Ross, 2003, chap. 12).  A $\chi^2$ test of 
independence of the contingency table from a product of marginals is used
with degrees of freedom associated with contingency table 
row and column dimensions: this provides a measure of how 
much structure we have, but not between alternative contingency
tables.  The latter is very fittingly addressed with the $\chi^2$
metric (see Murtagh, 2005) used in  correspondence analysis: we can 
say that correspondence analysis is the transformation of pairwise
$\chi^2$ distances into Euclidean distances, and that the latter greatly 
facilitates visualization (e.g., low-dimensional projection) and 
interpretation.   
The total inertia or trace of the data table grows 
with contingency table dimensionality, so that is of no direct help 
to us.  For the futures data used below,
and contingency 
tables of size $3 \times 3$, $4 \times 4$, $5 \times 5$, 
$6 \times 6$, and $10 \times 10$, we find traces of value: 
0.0118, 0.0268, 0.0275, 0.0493, and 0.0681, respectively.  Barring 
the presence of low-dimensional patterns arising in such a sequence of 
contingency tables, we will {\em always} find that greater 
dimensionality implies greater complexity (quantified, e.g., by trace) 
and therefore structure.


To address the issue of number of coding states to use, in order to search
for latent structure in such data, one approach that seems very 
reasonable is
to explore the dependencies and associations based on fine-grained
structure; and 
include in this exploration the possible aggregation of the 
fine-grained states.  (Aggregation of states 
 in correspondence analysis is catered for through the property of 
distributional equivalence: see Murtagh, 2005, for discussion.)

\subsection{Granularity of Coding}

We take sets of 
2500 values from the time series.  Tables \ref{table1} 
shows data to be analyzed derived from time series values 1 to 2500 
(identifier $i$).  Further, we use similar cross-tabulations for 
values 3001 to 5500 (identifier $k$), 2001 to 4500 
(identifier $m$), and values 3600 to 6100 (identifier $n$).  

\begin{table}
\caption{Cross-tabulation of log-differenced futures data using 
quantile coding with 10 current and next step price movements.
Values 1 to 2500 in the time series are used.  
Cross-tabulation results are expressed as percentage (by row).}
\label{table1}
\begin{verbatim}
       j1    j2    j3    j4    j5    j6    j7    j8    j9   j10
 i1  23.29  7.23  8.84  6.02 14.86  1.20 10.44  8.84  8.43 10.84 
 i2  11.60 11.60 11.20  8.80 13.20  5.20 11.60  8.80  8.80  9.20
 i3  10.00 13.20 10.80 12.80 14.40  2.00 12.80  5.60 10.80  7.60 
 i4   8.00  9.20  9.20 12.00 15.60  4.80 12.00 10.40  9.60  9.20
 i5   7.50  9.50  9.75 11.00 22.25  5.25  7.50 10.25  9.00  8.00 
 i6   5.05  8.08  9.09 10.10 20.20  6.06  9.09 16.16  4.04 12.12 
 i7   4.80  9.60 12.40 11.60 21.60  2.40 10.40  9.20 10.40  7.60 
 i8   8.40  7.20  8.40 12.40 13.20  7.20  8.40 10.80 11.60 12.40 
 i9   8.40 12.00  8.40  6.80 15.60  2.00 10.00 13.60  9.60 13.60 
i10  11.20 11.60 11.60  8.00  8.00  4.00  8.80 10.00 14.80 12.00
\end{verbatim}
\end{table}

\begin{figure}
\centerline{\vbox{\psfig{figure=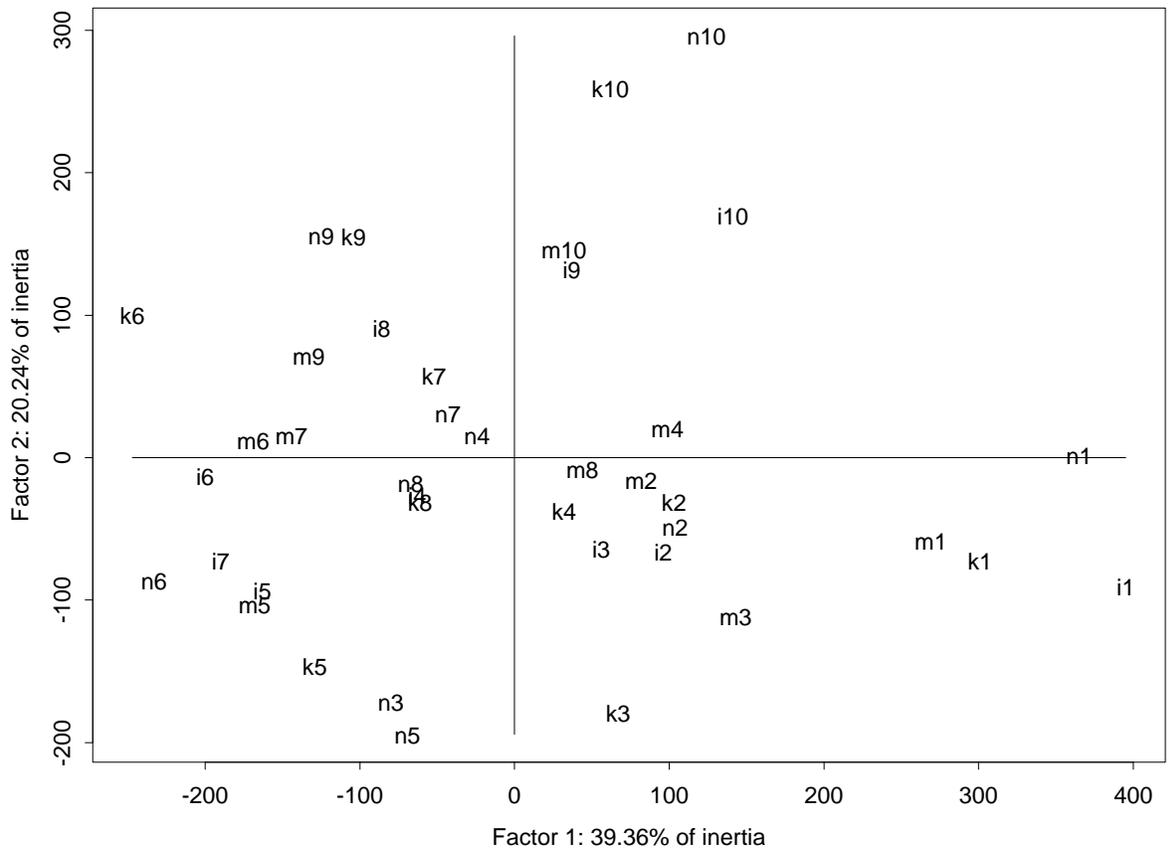,angle=270,width=16cm}}}
\caption{Factors 1 and 2 with input code categories 1 through 10
defined on 4 different spanning segments of the input data signal.
Only input, or current, values are displayed here.
The 4 time series sub-intervals 
are represented by (in sequential order) $i$, $m$, $k$, 
$n$.  The quantile coding is carried out independently in 
each set of 10 categories.}
\label{plot1}
\end{figure}

Figure \ref{plot1} shows the projections of the profiles in the plane
of factors 1 and 2, using all four data tables -- one of which is
shown  in Table 
\ref{table1}.  The result is very consistent: 
cf.\ how $\{ i1, k1, m1, n1 \}$ are tightly grouped, as are 
$\{ i2, k2, m2, n2 \}$, reasonably so $\{ i10, k10, m10, n10\}$, and so on. 
The full space of all factors has to be used to verify the clustering 
seen in this planar (least squares optimal) projection.  

An analysis of clusters found is listed in Table \ref{table5}.  
(Contributions to, and correlations with, the principal factors are
used: see Murtagh, 2005, for a discussion of where these may differ
from projections onto the factors.
Projections, e.g.\ as shown in Figure \ref{plot1}, are descriptive: 
``what is?'',  but correlations and contributions 
point to influence: ``what causes?''.  
Correlations and contributions are 
used therefore, in preference to projections.) 

In cluster 65, coding category 9 is predominant.  In cluster 68, 
coding categories 2 and 3 are predominant.  Cluster 69 is mixed. 
Cluster 70 is dominated by coding category 10.  In cluster 71, 
coding category 8 is predominant.  Cluster 72 is defined by coding category 
1.  Finally, cluster 73 is dominated by coding category 5.

\begin{table}
\caption{Table crossing clusters (on $I$) and coordinates ($J$), giving
correlations and contributions (as thousandths).  Clusters are labeled:
65, 68, 69, 70, 71, 72, 73.}  
\label{table5}
{\small
\begin{verbatim}

            Clusters                       Quantile coding category
        
Cluster 65: k9 n9 k7 n7 i4 m9              Predominant: 9 
Cluster 68: i3 k3 m3 m4 i2 m2 k2 n2        Predominant: 2, 3
Cluster 69: n6 i8 m7                       Predominant: none
Cluster 70: i10 m10 i9 k10 n10             Predominant: 10
Cluster 71: i6 k4 n4 m8 k8 n8              Predominant: 8
Cluster 72: i1 m1 k1 n1                    Predominant: 1
Cluster 73: i5 m5 n3 k5 n5 k6 i7 m6        Predominant: 5

\end{verbatim}
}
\end{table}

From the clustering, we provisionally retain coding categories 
1; 2 and 3 together; 5; 8; 9; and 10.  We flag response categories 
4, 6, and 7 as being unclear and best avoided when our aim is prediction 
of the futures data.  

To check the coding relative to stationarity, 
we check 
that the global code boundaries are close to the time series 
sub-interval code boundaries.  (See Murtagh, 2005, for more discussion 
on this, including confirmation of stationarity.)  In broad terms, what 
we are checking here is the consistency of the representative elements,
found in different subsets of the data, as illustrated above, right 
at the start of our presentation in this article, in Table 
\ref{pic1}.  


\section{Fingerprinting the Price Movements}

Typical movements can be read off in percentage terms in a table such as 
Table 
\ref{table1}.  More atypical movements serve to define the 
strong patterns in our data.  

We consider the clusters of current time-step code categories numbered 
65, 68, 69, 70, 71, 72, 73 from Table \ref{table5}, 
and we ask what are the likely movements,
for one time step.  Alternatively expressed the current code
categories are defined at time step $t$, and the one-step-ahead code
categories are defined at time step $t+1$.  

We find the following predominant movements in Table \ref{table5}
(using a thresholded contribution value -- not shown here; we recall that
``contribution'' is used in the correspondence analysis sense, meaning
mass times projection squared):

Cluster 65, i.e.\ code category 9:  $\longrightarrow$ weakly 8 and more weakly
9.

Cluster 68, i.e.\ code categories 2 and 3: $\longrightarrow$ 7.

Cluster 69, i.e.\ mixed code categories: $\longrightarrow$ 6.

Cluster 70, i.e.\ code category 10: $\longrightarrow$ 10.

Cluster 71, i.e.\ code category 8:  $\longrightarrow$ weakly 8.

Cluster 72, i.e.\ code category 1: $\longrightarrow$ 1.

Cluster 73, i.e.\ code category 5: $\longrightarrow$ 5

\medskip

Consider the situation of using these results in an operational 
setting.  From informative structure, we have found that code 
category 1 (values less than the 10th percentile, i.e.\ very low) has a
tendency, departing from typical tendencies, to be prior to code category 1 
(again very low).  From any or all of tables such as Table \ref{table1}
we
can see how often we are likely to have this situation in practice: 
19.04\% (= average of 23.29\% from Table \ref{table1}, and 17.67\%, 
16.4\%, and 18.8\%, from the other analogous tables not shown here), 
given that we 
have code category 1.

Applying a similar fingerprinting analysis 
to  Ross's (2003) oil data, 749 values, we found that clustering the 
initial code categories did not make much sense: we retained therefore 
the trivial partition with all 10 code categories.  For the output
or one-step-ahead future 
code categories, we agglomerated 6 and 7, and denoted this cluster as 11.
We find the following, generally 
weak, associations derived from the contributions.

\medskip

Input code category 6  $\longrightarrow$ output code categories 1, 10 (weak).

Input code category 3  $\longrightarrow$ output code category 2.

Input code category 4  $\longrightarrow$ output code category 4.

Input code categories 9, 2  $\longrightarrow$ output code category 5 (weak).

Input code category 10  $\longrightarrow$ output code category 8.

Not surprisingly, 
we find very different patterns in the two data sets of different natures
used, the futures and the oil price signals. 

We have shown that structure can be discovered 
in data where such structure is 
not otherwise apparent.  Furthermore we have used correspondence analysis,
availing of its spatial projection and clustering aspects, 
as a convenient analysis environment.  Validating the conclusions 
drawn is always most important, and this is facilitated by (i) semi-interactive
data analysis, and (ii) consistency of results across subsets of the 
domain under investigation, $\Omega$.  

\section{Conclusions}

Our new framework for data, and the handling of data (including our defining
of a normed vector space), could be considered in a sense as ``only'' 
formalizing standard data analysis practice.  But in the exploration and 
analysis of complex phenomena (cf.\ the search for local structure and 
patterns in price movements) we need to be sure of our belief in how our
data express the underlying phenomena.  The traditional
Kolmogorov approach based on Lebesgue integration and sigma
algebras of probability-measurable sets is unnecessarily abstract and 
therefore largely
ignored by the ``engineering'' or pragmatic common sense of the data 
analyst.  

In this article we have shown how the generalized Riemann integral lends itself
to a more transparent definition of probability, in line with empirical 
data analysis practice.  As a foundation for our data analysis tasks, it
achieves a far better cohesiveness between data, and data analyses, vis 
\`a vis the underlying phenomena.  

\section*{Acknowledgements}
Some of this work was carried out in the project ``Integration 
Methods in Financial Analysis'', 
supported by the British Council, UK, and the Polish State 
Committee for Scientific Research, KBN, Poland.  




\section*{References}

\begin{enumerate}

\item 
J.P. Benz\'ecri, L'Analyse des Donn\'ees.  Tome II. 
L'Analyse des Correspondances, 2nd ed., Dunod, 1976.

\item 
J.P. Benz\'ecri, 
Correspondence Analysis Handbook, Marcel Dekker, 1992.


\item
R. Gordon, The Lebesgue, Denjoy, Perron and
Henstock Integrals, American Mathematical Society, 1994.

\item
R. Henstock, Lectures on the Theory of Integration,
World Scientific, Singapore, 1988.


\item
R.N. Mantegna and H.E. Stanley, An Introduction to 
Econophysics, Cambridge University Press, 2000.

\item 
P. Muldowney, A General Theory of Integration in
Function Spaces, Pitman Research Notes in Mathematics no. 153,
Harlow, 1988.

\item 
P. Muldowney, Topics in probability using
generalised Riemann integration, Mathematical Proceedings of the
Royal Irish Academy, 99(B)1 (1999) 39--50.


\item 
P. Muldowney, The Henstock integral and the
Black-Scholes theory of derivative asset pricing, Real Analysis
Exchange, 25(1) (2000/2001) 117--132


\item 
P. Muldowney and V.A. Skvortsov, Lebesgue
integrability implies generalized Riemann integrability in
$\R^{[0,1]}$, Real Analysis Exchange, 27(1) (2001/2002) 
223--234.





\item 
F. Murtagh, Correspondence Analysis and Data Coding with R and Java,
Chapman and Hall/CRC Press, 2005.

\item
S.M. Ross, An Elementary Introduction to Mathematical 
Finance, 2nd ed., Cambridge University Press, 2003.

\item 
W. Rudin, Real and Complex Analysis, McGraw-Hill,
New York, 1970.


\item
P.F. Velleman and L. Wilkinson, Nominal, ordinal, interval and ratio
typologies are misleading, The American Statistician, 47 (1993) 65--72.


\end{enumerate}

\end{document}